\documentclass[11pt]{article} 

\newcommand{\bea}{\begin{eqnarray}}
\newcommand{\eea}{\end{eqnarray}}
\newcommand{\lsim}{{\;\raise0.3ex\hbox{$<$\kern-0.75em\raise-1.1ex\hbox{$\sim$}}\;}}
\newcommand{\gsim}{{\;\raise0.3ex\hbox{$>$\kern-0.75em\raise-1.1ex\hbox{$\sim$}}\;}}
 
\begin{document}

\begin{flushright}
{\large HIP-2002-58/TH}\\
\end{flushright}
\vspace{0.1cm}

\begin{center}
{\LARGE\bf Upper bounds on the mass of the lightest neutralino}
\\[15mm]
{\bf K. Huitu$^{a,b}$, J. Laamanen$^b$, and P.N. Pandita$^c$}\\[4mm]
$^a$High Energy Physics Division, Department of Physical Sciences, 
University of Helsinki, and
$^b$Helsinki Institute of Physics, 
P.O.Box 64, FIN-00014 University of  Helsinki, Finland \\
$^c$Department of Physics, North Eastern Hill University,
Shillong 793 022, India\\[7mm]
\date{}
\end{center}

\begin{abstract}

We derive the general upper bounds on the mass of the lightest
neutralino, as a function of the gluino mass, in different
supersymmetry breaking models with minimal particle content and the
standard model gauge group.  This includes models with gravity
mediated supersymmetry breaking~(SUGRA), as well as models with
anomaly mediated supersymmetry breaking~(AMSB).  We include the
next-to-leading order corrections in our evaluation of these bounds.
We then expand the mass matrix in powers of $M_Z/\mu$, and find the
upper bound on the mass of the lightest neutralino from this
expansion.  By scanning over all of the parameter space, we find that
the bound we have obtained can be saturated.  We compare the general
upper bound on the lightest neutralino mass to the upper bound that is
obtained when the radiative electroweak symmetry breaking scenario is
assumed.

\end{abstract}
\vskip 7mm

{~~~~PACS number(s): 12.60.Jv, 14.80.Ly}
\section{Introduction} 

It is widely expected that at least some supersymmetric particles will
be produced at the large hadron collider (LHC) that is starting
operation in a few years' time.  However, most of these supersymmetric
particles will not be detected as such, since they will decay into the
particles of the Standard Model (SM), or to the lightest
supersymmetric particle (LSP), which is stable as long as the R-parity
is conserved.  Thus, the experimental study of supersymmetry involves
the study of cascade decays of the supersymmetric particles to the LSP
and the reconstruction of the subsequent decay chains.  The LSP in a
large class of supersymmetry breaking models is the lightest
neutralino, which has thus been a subject of intense study for a long
time \cite{Choi:2001ww,pnp,neutgen,neutdec,neutDM,Berezinsky:1995cj}.
A stable lightest neutralino is also an excellent candidate for dark
matter~\cite{neutDM}.  As such it is important to have information on
the mass of the lightest neutralino state.

In view of this, the properties of the lightest neutralino and also
heavier neutralinos and charginos, which often appear in the cascade
decays, are of considerable importance.  In the minimal
version\footnote {By minimal version we here mean the model with the
minimal particle content and the Standard Model gauge group} of the
supersymmetric extension of the Standard Model at least two Higgs
doublets $H_1$ and $H_2$ with hypercharge~$({Y})$ having values $ -1 $
and $+1$, respectively, are required.  The fermionic partners of these
Higgs doublets mix with the fermionic partners of the gauge bosons to
produce four neutralino states $\tilde\chi^0_i, i=1,2,3,4$, and two
chargino states $\tilde\chi^{\pm}_i, i=1,2$.

The neutralino mass matrix $\hat \mathcal M$ depends on the ratio of
the vacuum expectation values (VEVs) of the two Higgs doublets denoted
by $\tan\beta \equiv v_2/v_1$, where $v_1 = \langle H_1^0\rangle$ and
$v_2 = \langle H_2^0\rangle$ are the vacuum expectation values of the
two Higgs doublets with opposite hypercharge, the supersymmetry
breaking $U(1)_Y$ and $SU(2)_L$ gaugino masses $M_1$ and $M_2$, and
the supersymmetry conserving Higgs(ino) mixing parameter $\mu$.  The
mass matrix is symmetric, but not necessarily real.  The mass
parameters can have arbitrary complex phases, as can also the Higgs
boson VEVs. However, all of these are not actually independent -- one
can choose the two nontrivial phases to be in $M_1$ and $\mu$.  The
electric dipole moments strictly constrain the phases in
supersymmetric (SUSY) models.  However, these bounds are for products
of the phases.  Thus, if there are cancellations between phases, a
single phase can be larger than the limits for the product \cite{prs}.

Recently, it has been demonstrated that at the linear collider one can
determine the above parameters of the neutralino and chargino sectors
from the masses of charginos and three lightest neutralinos, or
alternatively from two lightest neutralinos and the cross section
$e^+e^-\rightarrow \chi_1^0\chi_2^0$ \cite{Choi:2001ww,Choi:1998ei}.
The linear collider is likely to be available several years after the
completion of the LHC, and thus all the information that is available
now or can be obtained at the LHC will be very valuable.

In this paper we obtain the theoretical upper bound on the mass of the
lightest neutralino state in the most commonly studied supersymmetry
breaking models.  These include the gravity mediated supersymmetry
breaking model and the anomaly mediated supersymmetry breaking model,
with minimal particle content.\footnote{ In the gauge mediated
supersymmetry breaking (GMSB) models the lightest neutralino is not
the lightest supersymmetric particle.  However, in many models it is
the next-to-lightest particle.  Here we will also comment on the upper
bound in the GMSB models.}  In a general model with an arbitrary
particle content, an upper bound for the lightest neutralino mass was
calculated in \cite{pnp}.  For specific supersymmetry breaking
scenarios we can give a more accurate bound.  In Section 2 we obtain
the general upper bound on the mass of the lightest neutralino in the
minimal version of the supersymmetric standard model.  We then
evaluate this upper bound for the two most popular supersymmetry
breaking models, namely, the gravity mediated supersymmetry breaking
(SUGRA) models and the anomaly mediated supersymmetry breaking (AMSB)
models. We include the next-to-leading order corrections in the
numerical evaluation of this upper bound.

In Section 3, we then study the expansion of the neutralino mass
matrix in powers of $M_Z/\mu$, which at second order is accurate to
1\% for large values of $\mu$.  The extremum (maximum) of the lightest
neutralino mass gives an upper bound from this expansion.  This value
is lower than the upper bound obtained directly from the mass matrix.
By a numerical scan over real and complex parameter values we confirm
that this bound is accurately saturated in the supersymmetric models
that we study in this paper.  We compare the general upper bound with
the largest value of the lightest neutralino mass that is obtained
when radiative electroweak symmetry breaking is assumed (and the $\mu$
parameter determined from the radiative electroweak symmetry
breaking).  For this purpose we have used the numerical program
SOFTSUSY~\cite{Allanach:2001kg}. In Section 4 we present our
conclusions.

\section{The general upper bound on the mass of the lightest neutralino}

We start by recalling the neutralino mass matrix in supersymmetric
models in the basis \bea \psi^0_j = (-i\lambda',~ -i\lambda_3,~
\psi^1_{H_1},~ \psi^2_{H_2}),~~~ j = 1,~2,~3,~4,
\label{neut1}
\eea which can be written as~\cite{nilles} \bea \hat {\mathcal
M}=\left(
\begin{array}{cccc} M_1 & 0 & -M_Z\cos\beta\sin\theta_W & 
M_Z\sin\beta\sin\theta_W\\
0 & M_2 & M_Z\cos\beta\cos\theta_W & -M_Z\sin\beta\cos\theta_W\\ 
-M_Z\cos\beta\sin\theta_W& M_Z\cos\beta\cos\theta_W & 0 &-\mu\\
M_Z\sin\beta\sin\theta_W & -M_Z\sin\beta\cos\theta_W & -\mu & 0\\
\end{array} \right), \label{neutmatrix}
\eea where $\lambda'$ and $\lambda_3$ are the two-component gaugino
states corresponding to the $U(1)_Y$ and the third component of the
$SU(2)_L$ gauge groups, respectively, and $\psi^1_{H_1}, \psi^2_{H_2}$
are the two-component Higgsino states.  Furthermore, $g'$ and $g$ are
the gauge couplings associated with the $U(1)_Y$ and the $SU(2)_L$
gauge groups, respectively, with $\tan\theta_W = g'/g$, and $M_Z^2 =
(g^2 +g'^2)(v_1^2 + v_2^2)/2.$ Assuming CP conservation, this mass
matrix is real.  We shall denote the eigenstates of the neutralino
mass matrix by $\chi^0_1, \chi^0_2, \chi^0_3, \chi^0_4$ labeled in
order of increasing mass. Since some of the neutralino masses
resulting from diagonalization of the mass matrix can be negative, we
shall consider the squared mass matrix $ \hat {\mathcal
M}^{\dagger}\hat {\mathcal M}$.  An upper bound on the squared mass of
the lightest neutralino $\chi^0_1$ can be obtained by using the fact
that the smallest eigenvalue of $ \hat {\mathcal M}^{\dagger}\hat
{\mathcal M}$ is smaller than the smallest eigenvalue of its upper
left $2 \times 2$ submatrix

\bigskip
\begin{equation}
\left(
\begin{array}{lr}
M_1^2 + M_Z^2\sin^2\theta_W & -M_Z^2\sin\theta_W \cos\theta_W\\
&\\
-M_Z^2\sin\theta_W \cos\theta_W & M_2^2 + M_Z^2\cos^2\theta_W
\end{array}
\right),
\label{submatrix1} 
\end{equation}

\bigskip
\noindent
thereby resulting in the upper bound
\bea
M_{\chi^0_1}^2  \le  \frac 12 \left(M_1^2 + M_2^2 +M_Z^2
- \sqrt{(M_1^2 - M_2^2)^2 +M_Z^4 -2 (M_1^2 - M_2^2)M_Z^2\cos
2\theta_W }\right).
\label{bound1}
\eea 
We emphasize that the upper bound (\ref{bound1}) is independent of the
supersymmetry conserving parameter $\mu$ and also independent of
$\tan\beta$, but depends on the supersymmetry breaking gaugino mass
parameters $M_1$ and $M_2$. Despite this dependence on the unknown
supersymmetry breaking parameters, we will show that
Eq.~(\ref{bound1}) leads to a useful bound on $M_{\chi_1^0}$.

\subsection{Gravity mediated supersymmetry breaking}

In the gravity mediated minimal supersymmetric standard model, the
soft gaugino masses $M_i$ satisfy the renormalization group
equations~(RGEs)~($|M_3| = m_{\tilde g}$, the gluino mass) \bea
16\pi^2\frac{dM_i}{dt} = 2 b_i M_i g_i^2, ~~~~b_i =
\left(\frac{33}{5}, 1, -3\right)
\label{gaugino1}
\eea
at the leading order. Here $g_1 =\frac{5}{3}g',\; g_2 = g$, and $g_3$ 
is the $SU(3)_C$ gauge coupling. The RGEs (\ref{gaugino1}) imply that
the soft supersymmetry breaking gaugino masses scale like
gauge couplings:
\bea
\frac{M_1(M_Z)}{\alpha_1(M_Z)} = \frac{M_2(M_Z)}{\alpha_2(M_Z)}
= \frac{M_3(M_Z)}{\alpha_3(M_Z)} ,
\label{gaugino2}
\eea
where $\alpha_i = g_i^2/4\pi, i = 1, 2, 3$. 

The relation (\ref{gaugino2}) reduces the three gaugino mass parameters to
one, which we take to be the gluino mass $m_{\tilde g}$. The other
gaugino mass parameters are then determined through 
\bea M_1(M_Z) & =
& \frac{5 \alpha}{3 \alpha_3~\cos^2\theta_W}~m_{\tilde g} ~~\simeq~~
0.14~m_{\tilde g},
\label{m3relation}\\
M_2(M_Z) & = & \frac{\alpha}{\alpha_3~\sin^2\theta_W}~m_{\tilde g}
~~\simeq~~ 0.28~m_{\tilde g} ,
\label{m2relation}
\eea where we have used the value of various couplings at the $Z^0$
mass \bea \alpha^{-1}(M_Z) = 127.9, ~~~~~ \sin^2\theta_W = 0.23, ~~~~~
\alpha_3(M_Z) = 0.12.  \eea 
Using Eqs.~(\ref{m3relation}) and (\ref{m2relation}) in
Eq.~(\ref{bound1}), we get the upper bound on the mass of the lightest
neutralino.\footnote{ In the GMSB models, one gets the same relations
(\ref{gaugino2}) at the messenger scale, since $M_a \propto \alpha_a$.
Thus, the upper bound obtained in the SUGRA model can be applied in
the GMSB model as well.}  For a gluino mass of 200 GeV, the upper
bound (\ref{bound1}) for the lightest neutralino mass is about
$35$~GeV. Similarly, for a gluino mass of $1$ TeV, the upper bound
(\ref{bound1}) becomes $186$ GeV.

\begin{figure}[t]
\leavevmode
  \begin{center}
    \mbox{\epsfxsize=11truecm \epsfysize=7truecm \epsffile{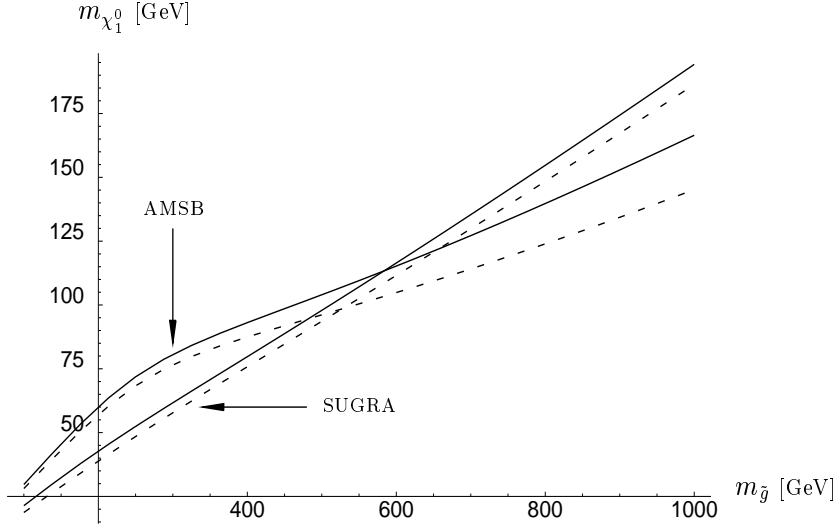}}
  \end{center}
  \caption{The upper bound on the mass of the lightest neutralino in
the SUGRA and AMSB models. The tree level results are given by dashed
lines and the next-to-leading order results by the solid lines.}
\label{twobytwo}
\end{figure}

We have plotted the upper bound (\ref{bound1}) on the mass of
the lightest neutralino in  Fig.~\ref{twobytwo}.
The almost straight dashed line corresponds to the SUGRA model at the
tree level.
From  Fig.~\ref{twobytwo}, we observe  
that $m_{\chi^0_1} < 186$ GeV for $m_{\tilde g} < 1$ TeV.
\input{epsf.sty}

We now include next-to-leading order (NLO) corrections coming from
$\alpha_3$ and from the top-quark Yukawa coupling $\alpha_t\ (\equiv
h_t^2/4\pi)$ two-loop contributions to the beta functions and
logarithmically enhanced weak threshold corrections.  In this
approximation, one finds \cite{ggw}
\bea
M_1^{NLO}&=& M_1(Q)\left\{ 1+\frac{\alpha}{8\pi \cos^2\theta_W}
\left[ -21 \ln \frac{Q^2}{M_1^2} +11 \ln \frac{m_{\tilde q}^2}{M_1^2}
+9\ln \frac{m_{\tilde \ell}^2}{M_1^2}
\right. \right. \nonumber \\
&&+ \left. \left. \ln \frac{\mu^2}{M_1^2}
+\frac{2\mu}{M_1}\sin 2\beta \frac{m_A^2}{\mu^2-m_A^2}
\ln \frac{\mu^2}{m_A^2}\right] +\frac{2\alpha_3}{3\pi} 
-\frac{13\alpha_t}{66\pi}\right\}, \label{gnlo1}
\eea
\bea
M_2^{NLO}&=& M_2(Q)\left\{ 1+\frac{\alpha}{8\pi \sin^2\theta_W}
\left[ -13 \ln \frac{Q^2}{M_2^2} +9 \ln \frac{m_{\tilde q}^2}{M_2^2}
+3\ln \frac{m_{\tilde \ell}^2}{M_2^2}
\right. \right. \nonumber \\
&&+ \left. \left. \ln \frac{\mu^2}{M_2^2}
 +\frac{2\mu}{M_2}\sin 2\beta \frac{m_A^2}{\mu^2-m_A^2}
\ln \frac{\mu^2}{m_A^2}\right] +\frac{6\alpha_3}{\pi} -\frac{3\alpha_t}{2\pi}
\label{gnlo2}
\right\} ,
\eea
\bea
M_3^{NLO}&=& M_3(Q)\left\{ 1+\frac{3\alpha_3}{4\pi}
\left[ \ln \frac{Q^2}{M_3^2} + F\left( \frac{m_{\tilde q}^2}{M_3^2}\right)
-\frac{14}{9}\right] +\frac{\alpha_t}{3\pi} \right\}, \\ \label{gnlo3}
F(x)&=& 1+2x+2x(2-x)\ln x +2(1-x)^2\ln |1-x| \label{aux1}.
\eea
Here $M_i(Q)$ are the leading order results given by
Eq.~(\ref{gaugino2}).  Notice that the next-to-leading order
corrections are of the same form in all models.  It is only the
leading order $M_i(Q)$ that are different for different models.

In order to calculate the next-to-leading order upper bound on the
lightest neutralino mass, we need to know $M_1^{NLO}$ and
$M_2^{NLO}$. As with the leading order result, we express $M_1^{NLO}$
and $M_2^{NLO}$ as a function of $M_3^{NLO}$ (the NLO physical gluino
mass), using Eqs.~(\ref{gnlo1}) -- (\ref{gnlo3}) and substitute it in
Eq.~(\ref{bound1}).  We plot the NLO corrected upper bound on the
lightest neutralino mass as a function of the gluino mass as a solid
curve in Fig.~\ref{twobytwo}.  As input we have used here $\tan\beta =
10$, $m_t(pole) = 174$ GeV, $m_0=300$ GeV, $A_0=1$ TeV, $\mu=-460$
GeV, and $Q=890$ GeV.  Since dependence on these model parameters
appears only at the loop level, the upper bound is not very sensitive
to these parameters.  We note that the NLO corrections increase the
upper bound from its tree level result by only a few GeV for a wide
range of the gluino mass.  Indeed, we find that at the NLO the upper
bound on the mass of the lightest neutralino is $m_{\chi^0_1} < 194$
GeV for $m_{\tilde g} < 1$ TeV.  In \cite{Pierce:1996zz} full one-loop
corrections to sparticle masses were calculated.  The loop corrections
to the lightest neutralino mass can be typically 10\% for
$m_{\chi^0_1} < 40$ GeV.  However, for $m_{\chi^0_1} =100$ GeV the
full one loop corrections to the lightest neutralino mass are $\lsim
5$\%.  We note that the experimental lower bound~\cite{pdg} on the
lightest neutralino mass, valid for any $\tan\beta$ and $m_0$, is
$m_{\chi_1^0}>37$ GeV.

\subsection{The anomaly mediated supersymmetry breaking}

The anomaly induced soft terms are always present in a broken
supergravity theory, regardless of the specific form of the couplings
between the hidden and observable sectors. They are linked to the
existence of the superconformal anomaly. Indeed, they explicitly arise
when one tries to eliminate from the relevant Lagrangian the
supersymmetry breaking auxiliary background field by making a suitable
Weyl rescaling of the superfields in the observable sector.

The soft terms in the anomaly mediated supersymmetry breaking
models are especially interesting because they are invariant under the
renormalization group transformations.  The phenomenological
appeal of the soft terms in AMSB resides precisely in this crucial
property. In particular, it implies a large degree of predictivity,
since all the soft terms can be computed from the known low-energy SM
parameters and a single mass scale $m_{3/2}$. Also, it leads to robust
predictions, since the RG invariance guarantees complete insensitivity
of the soft terms to the ultraviolet physics. As demonstrated with
specific examples in Ref.~\cite{giudice}, heavy states do not affect
the low-energy parameters, since their effects in the beta functions
and threshold corrections exactly compensate each other.  This means
that the gaugino mass prediction
\begin{eqnarray}
\label{spectroscopy}
     M_\lambda &=& {\beta_g\over g} m_{3/2}
\end{eqnarray}
is valid irrespective of the grand unified theory gauge group in which
the SM may or may not be embedded.  A unique feature of the
anomaly mediated supersymmetry is the gaugino mass hierarchy implied
by Eq.~(\ref{spectroscopy}).  At the leading order, we  thus  have
\bea
M_1(Q)&=&\frac{11\alpha (Q)}{4\pi \cos^2\theta_W} m_{3/2}
\simeq  8.9 \times 10^{-3}  m_{3/2},  \label{lead1}\\
M_2(Q)&=&\frac{\alpha (Q)}{4\pi \sin^2\theta_W} m_{3/2}
\simeq 2.7 \times 10^{-3} m_{3/2},  \label{lead2}\\
M_3(Q)&=&-\frac{3\alpha_3 (Q)}{4\pi} m_{3/2}
\simeq - 2.8 \times 10^{-2} m_{3/2}, \label{lead3}
\eea
at the scale $M_Z$.  Using Eqs.~(\ref{lead1}) -- (\ref{lead3}) in
Eq.~(\ref{bound1}), we obtain the leading order result for the upper bound
on the lightest neutralino mass in the minimal AMSB model.  We have
plotted this upper limit as the upper dashed curve in the
Fig.~\ref{twobytwo}.  It is interesting to note that there is a kink
in this dashed curve around $m_{\tilde g}\simeq 210$ GeV.  This is due
to the competition between the diagonal terms in the $2\times 2$
submatrix~(\ref{submatrix1}).
The term containing $M_1$ is smaller, when the gluino mass is small,
but with the increasing gluino mass the term with $M_2$ becomes
smaller around 210 GeV.  This is because the Wino triplet mass
parameter is always smaller than the Bino mass parameter in the AMSB
type model, contrary to the SUGRA or GMSB type models where the Bino
mass parameter is smaller than $M_2$.

In the next-to-leading order corrections to the lightest neutralino
mass in AMSB models, the complete  sparticle spectrum becomes important.
Unfortunately, it turns out that the pure scalar mass squared anomaly
contribution for the sleptons is negative~\cite{randall}.  In order to
avoid this problem we need to consider other positive soft
contributions to the spectrum.  This can arise in a number of ways,
but most of the solutions will spoil the RG invariance of the soft
terms and the consequent ultraviolet insensitivity.  Nevertheless,
there are various options to cure this problem without reintroducing
the flavor problem \cite{randall,pr,fix, hlp}.

The necessary cure for the slepton masses may also completely upset 
the mass relations for the other particles (as in the case of the
model of Ref.~\cite{pr}).  However, here we will simply parameterize
the new positive contributions to the squared sfermion masses with a
common mass parameter $m_0^2$, assuming that the extra terms do not
reintroduce the supersymmetric flavor problem.  The low-energy soft
supersymmetry breaking parameters for the scalars and the trilinear
couplings are then obtained from
\begin{eqnarray}
     m_{\tilde Q}^2&=&-{1\over 4}
      \left({\partial\gamma\over\partial g}\beta_g +
       {\partial\gamma\over\partial y}\beta_y\right) m_{3/2}^2 +m_0^2,
\label{budda} \\
\label{spectroscopy3}
     A_{y}&=&-{\beta_y\over y} m_{3/2},
\end{eqnarray}

\noindent
respectively. Using Eqs.~(\ref{gnlo1}) --  (\ref{gnlo3}), 
we obtain for the anomaly mediated supersymmetry breaking models the 
next-to-leading order results for the gaugino mass parameters as 
\bea
M_1^{NLO}&=& 1.06\, M_1(Q),  \label{nlo1} \\
M_2^{NLO}&=& 1.28\, M_2(Q),  \label{nlo2} \\
M_3^{NLO}&=&0.9\, M_3(Q),  \label{nlo3}
\eea
where the $M_i(Q),  i = 1, 2, 3$ (the leading order result), 
is given in Eqs.~(\ref{lead1})--(\ref{lead3}).
Here we have used as input 
$\tan\beta  =  10$,
$m_t(pole) = 174$ GeV,
$m_{3/2}  = 35$ TeV,
$m_0=600$ GeV, 
$\mu=-600$ GeV, and $Q=958$ GeV.
The Higgsino corrections to $M_1$ and $M_2$ are proportional to
$\mu / M_{1,2}$ and can become very important in models with large $\mu$,
as discussed in Ref.~\cite{giudice}.

In Fig.~\ref{twobytwo} we have plotted the next-to-leading order upper
bound on the mass of the lightest neutralino in anomaly mediated
supersymmetry breaking models.  The NLO result, obtained  using
Eqs.~(\ref{nlo1}) -- (\ref{nlo3}), is shown as a solid line.  The NLO
corrections are significant, of the order of 20\%.  The larger NLO
correction in the AMSB model as compared to the SUGRA model is due to
the fact that the $\alpha_3$ corrections for the $M_2$ mass parameter
are larger than for the $M_1$ parameter.  For $m_{\tilde g}<1$ TeV,
the upper bound on the lightest gluino mass is 167 GeV, which is
considerably less than in the SUGRA case.

\section{Lightest neutralino mass bound from the structure 
of the mass matrix}

We can also obtain information on the neutralino masses by studying the 
expansion of the neutralino mass matrix in terms of the parameter $M_Z/\mu$.
This  expansion can be obtained  most conveniently  by using the  basis
$(\tilde\gamma, \tilde Z^0, \tilde H^0_a,\tilde H^0_b)$. In this basis
the mass matrix is given by 
\bea 
\hat {\mathcal M}=\left(
\begin{array}{cccc} M_1 c^2_W +M_2 s^2_W & (M_2- M_1)c_W s_W & 0 & 0
\\ (M_2- M_1)c_W s_W & M_1 s^2_W +M_2 c^2_W & M_Z &0\\ 0&M_Z & \mu
s_{2\beta} &-\mu c_{2\beta}\\ 0&0&-\mu c_{2\beta}& \mu s_{2\beta}\\
\end{array} \right). \label{matrixGB} 
\eea 
Here we have used the abbreviations $s_{2\beta} =\sin 2\beta$,
$c_{2\beta} =\cos 2\beta$, $s^2_W =\sin^2 \theta_W$ and $c^2_W =\cos^2
\theta_W$.  Let us start by supposing, as before, that all the mass
parameters are real.  The mass matrix is then real and
symmetric.\footnote {We note that in the specific models that we have
been considering, SUGRA and AMSB, the phases of $M_1$ and $M_2$ are
the same (see Eqs.~(\ref{gaugino2}) and (\ref{spectroscopy})). So if
$M_2$ is real, then $M_1$ is also real. On the other hand, the $\mu$
parameter is in general complex.  Complex parameters would imply a
non-Hermitian mass matrix, giving generally complex eigenvalues.  Such
a situation can be handled by considering the eigenvalues of the
matrix $\hat {\mathcal M}^\dagger\hat {\mathcal M}$.}  The neutralino
mass matrix $\hat {\mathcal M}$ can be cast into a form whereby the
gaugino and Higgsino mass parameters are only at the diagonal
positions by a similarity transformation with a matrix ${\mathcal A}$,
\bea {\mathcal M}= {\mathcal A}^T \hat{\mathcal M}
{\mathcal A}, \eea where \bea {\mathcal A}=\left(
\begin{array}{cccc}  c_W  &  s_W & 0 & 0
\\ - s_W &  c_W & 0 &0\\ 0& 0 & \cos(\pi /4-\beta) &\sin(\pi/4 -\beta)
\\ 0&0&-\sin(\pi/4 -\beta)& \cos (\pi/4-\beta)\\
\end{array} \right). \label{matrixA} 
\eea 
The mass matrix can then be diagonalized by using perturbation theory.
In the SUGRA model, for the mass of the lightest
neutralino we get, up to terms of $\mathcal O (M_Z/\mu)^2$, 
\bea m_{\chi^0_1}=M_1 - \frac{M_Z^2 s^2_W}\mu \sin 2\beta -
\left(M_Z^2 s^2_W M_1 +\frac{M_Z^4 s^2_W c^2_W}{M_2 -M_1}\sin ^2
2\beta \right)\frac 1 {\mu^2}. \label{lightestN} \eea Similarly, for
the second lightest neutralino ${\chi^0_2}$ we obtain \bea
m_{\chi^0_{2}}=M_2 - \frac{M_Z^2 c^2_W}\mu \sin 2\beta - \left(M_Z^2
c^2_W M_2 +\frac{M_Z^4 s^2_W c^2_W}{M_1-M_2 }\sin ^2 2\beta
\right)\frac 1 {\mu^2}. \label{2lightestN} \eea
If instead we were considering the
AMSB model, Eq.~(\ref{2lightestN}) would represent the mass of the lightest
neutralino $\chi^0_{1}$, and Eq.~(\ref{lightestN}) would give the formula
for the mass of the second lightest neutralino.  The dependence of the
lightest neutralino mass on the specific SUSY breaking scenario is due
to the fact that the ordering of the gaugino mass parameters is model
dependent (for AMSB models $M_2 < {M_1}$, whereas for SUGRA 
models ${M_1} < M_2$).

\begin{figure}
\leavevmode
  \begin{center}
    \mbox{\epsfxsize=10truecm \epsfysize=6.5truecm \epsffile{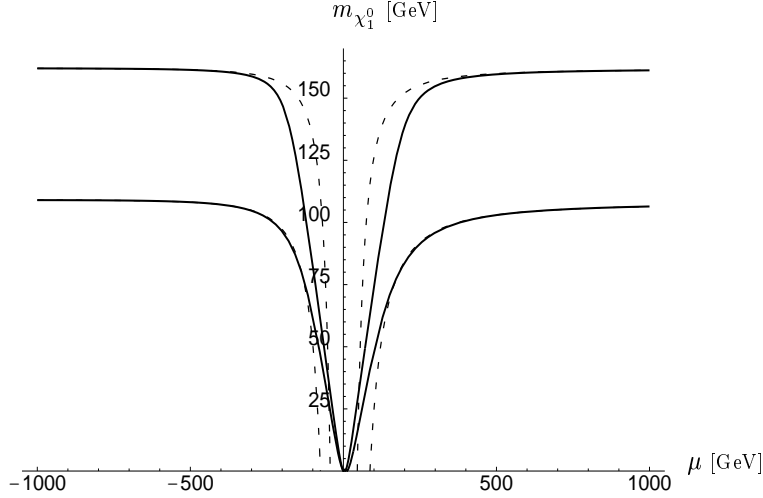}}
  \end{center}
  \caption{\label{hyvyys} Mass of the lightest neutralino $\chi^0_1$
    as a function of $\mu$. Solid lines correspond to the numerical
    results from the mass matrix and dashed lines to the second order
    expansion from Eq.~(\ref{lightestN}). The upper two lines
    represent masses in SUGRA models and the lower two in AMSB models.
    Here $\tan\beta=10$ and $m_{\tilde g}=900$ GeV.  }
\end{figure}

In Fig.~\ref{hyvyys} we plot the mass of the lightest neutralino
obtained from the expansion of the mass matrix in $(M_Z/\mu)$ together
with the exact results obtained from the numerical evaluation of the
lightest neutralino mass from the mass matrix.  The results for the
other neutralinos are very similar in accuracy.  The second order tree
level expansion is generally better than 1\% for $|\mu| > 450$ GeV
($m_{\tilde g}<1600$ GeV and $\tan \beta =10$), with the exception of
small gluino mass, when $m_{\chi^0_1}$ is very small, thus giving a
larger relative error. For our purpose it is sufficient to calculate
the expansion up to second order in $(M_Z/\mu)$.

\begin{figure}
\leavevmode
  \begin{center}
    \mbox{\epsfxsize=10truecm \epsfysize=6.5truecm \epsffile{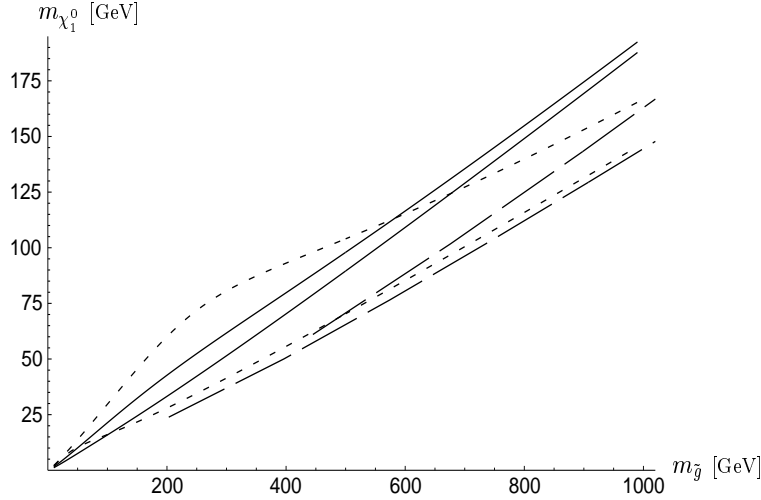}}
  \end{center}
  \caption{\label{uppercompare} The upper limit on $m_{\chi^0_1}$ as a
    function of $m_{\tilde g}$. Solid lines represent masses in SUGRA
    model and short dashed lines in AMSB model.  The lower curve in
    each case corresponds to the upper limit obtained from the
    expansion in $(M_Z/\mu)$, and the upper curve corresponds to the
    general upper limit obtained from the mass matrix.  The long
    dashed curves correspond to the case when the radiative
    electroweak symmetry breaking scenario is implemented.}
\end{figure}
Due to the simple functional form of Eqs.~(\ref{lightestN}) and
(\ref{2lightestN}) the extremal values of the masses with respect to
$\mu$ are easily calculated. These functions have only one extremum
(maximum), which is given (within the limits of validity of the
expansions) for the values of $\mu$
\bea \mu = -2\left( \frac{M}{\sin 2\beta} +\frac{M_Z^2 \sin
2\beta}{\tilde M -M}\frac{s^2_W c^2_W}{t} \right),
\label{myy}
\eea
and the maximum mass is then given by the upper bound
\bea
m_{\chi^0_1} \leq M + \frac 1 4 \frac {M_Z^2 t^2 \sin^2 2\beta}{
t M+\frac{M_Z^2 s^2_W c^2_W \sin^2 2\beta }{\tilde M -
M}},\label{chilimit} 
\eea 
where $M=\min({M_1},M_2), \tilde M=\max({M_1},M_2), t=s^2_W$ if
$M=M_1$ (SUGRA), and $t=c^2_W$ if $M=M_2$ (AMSB).  In
Fig.~\ref{uppercompare} we plot the upper limit on the lightest
neutralino mass obtained from Eq.~(\ref{chilimit}) as a function of
$m_{\tilde g}$, for both the SUGRA and AMSB models.  We also plot the
upper limit obtained from Eq.~(\ref{bound1}) in the same figure.  The
SUGRA results are represented as solid lines and the AMSB results as short
dashed lines.  The lower curve of each set corresponds to the upper
bound obtained from the expansion in $(M_Z/\mu)^2$, and the
upper curve corresponds to the upper bound obtained from 
Eq.~(\ref{bound1}).  These results for SUGRA and AMSB
in Fig.~\ref{uppercompare} are NLO results.  We have plotted the
results for the value of $\tan\beta =10$. In order to verify the accuracy
of these results,  we  made an extensive
scan over the parameter space, using both real and complex values of the
$\mu$ parameter. The highest mass  obtained  from this corresponds extremely
well to the upper limit obtained from the expansion in $(M_Z/\mu)$.

We have also made a scan over the parameter space using the SOFTSUSY
program \cite{Allanach:2001kg}, in which the phenomenon of radiative
electroweak symmetry breaking~(REWSB) is implemented.  Thus, the $\mu$
value in this program is given by the REWSB condition.  The resulting
spectrum includes one- and dominant two-loop corrections.  The maximum
mass obtained for the lightest neutralino is plotted in
Fig.~\ref{uppercompare} as a function of $m_{\tilde g}$ with
long-dashed lines.  The upper long-dashed line corresponds to the
SUGRA model and the lower one to the AMSB model.  One can see that
with radiative electroweak symmetry breaking, the $m_{\chi_1^0}$ in
the AMSB model is close to the maximum mass obtained from the
expansion in $(M_Z/\mu)$, while in the SUGRA model with REWSB the
$m_{\chi_1^0}$ obtained is clearly lower than the maximum value from
the expansion, indicating that $\mu_{REWSB}$ for the SUGRA model is
not close to the value obtained from Eq.~(\ref{myy}).

\begin{figure}
\leavevmode
  \begin{center}
    \mbox{\epsfxsize=10truecm \epsfysize=6.5truecm \epsffile{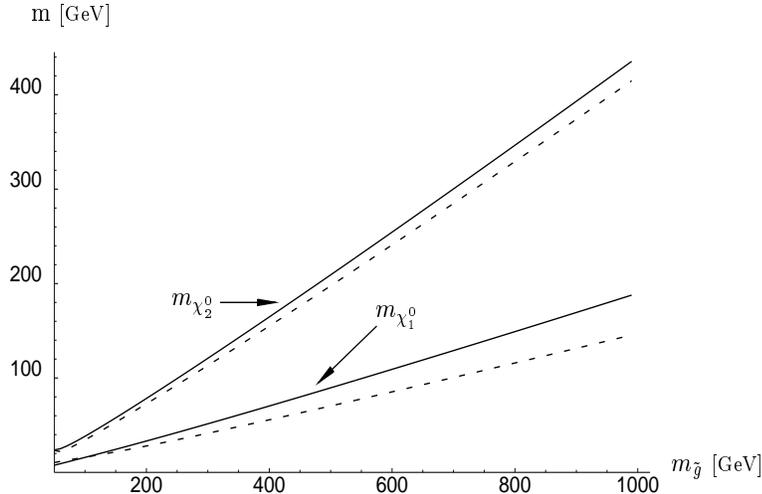}}
  \end{center}
  \caption{\label{n1etn2limits} The upper limits of $m_{\chi^0_1}$ and
    $m_{\chi^0_2}$ as a function of $m_{\tilde g}$.  Solid lines
    represent masses in SUGRA model and dashed in AMSB model.}
\end{figure}

As in the case for $\chi_1^0$ we can search for the upper bound on the
mass of the second lightest neutralino $\chi_2^0$.  For light gluinos
(lighter than ${\cal O}(60)$ GeV) the extremum in the mass for
$\chi_2^0$ is a minimum due to a sign change in the expansion, but for
experimentally allowed masses the extremum for $m_{\chi_2^0}$ is a
maximum.  In Fig.~\ref{n1etn2limits} we have plotted the upper limits
for both the lightest and second lightest neutralino obtained from the
expansion in $(M_Z/\mu)$.  The solid lines correspond to the SUGRA
model, while the dashed lines correspond to the AMSB model.  For
$m_{\tilde g}<1$ TeV, the NLO upper bounds for the second lightest
neutralino are 440 GeV for the SUGRA case and 419 GeV for the AMSB
case.

\section{Conclusions}

In this paper we have studied the neutralino mass matrix for the
minimal supersymmetric model with the aim of obtaining an upper bound
on the mass of lightest neutralino. Knowledge of the mass of the
lightest neutralino is of crucial importance for the supersymmetric
phenomenology.  We have shown that a general limit, valid for
arbitrary values of parameters, can be obtained from the mass matrix.
Even though such a bound depends on the supersymmetry breaking
parameters $M_1$ and $M_2$, it nevertheless leads to a significant
numerical bound on the lightest neutralino mass in the SUGRA and AMSB
models.  We have also obtained an upper bound on the lightest
neutralino mass by expanding the neutralino mass matrix in terms of
the parameter $M_Z/\mu$.  We see that the upper limit from this
expansion is considerably lower for the AMSB model than for the SUGRA
model for similar $m_{\tilde g}$.  From this analysis we conclude that
the upper bound on the mass of lightest neutralino is $ m_{\chi^0_1} <
200$ GeV for $m_{\tilde g} < 1$ TeV.

In Fig.~\ref{uppercompare} we have three separate regions for the
upper bound on the mass of the lightest neutralino: one which is valid
in both SUGRA and AMSB case, one which is valid only in one of the
models, and a third one which is not available for any of the models
that we have studied.

\begin{flushleft} {\bf Acknowledgments} \end{flushleft}

\noindent
KH and JL thank the Academy of Finland (project number 48787) for
financial support.  The work of PNP is supported by Department of
Science and Technology, India under project No. SP/S2/K-01/2000-II,
and by the Council of Scientific and Industrial Research, India.

\newcommand{\plb}[3]{{Phys. Lett.} {\bf B#1} #2 (#3)} %
\newcommand{\prl}[3]{Phys. Rev. Lett. {\bf #1} #2 (#3) } %
\newcommand{\rmp}[3]{Rev. Mod.  Phys. {\bf #1} #2 (#3)} %
\newcommand{\prep}[3]{Phys. Rep. {\bf #1} #2 (#3)}                   %
\newcommand{\rpp}[3]{Rep. Prog. Phys. {\bf #1} #2 (#3)}             %
\newcommand{\prd}[3]{Phys. Rev. {\bf D#1} #2 (#3)}                    %
\newcommand{\np}[3]{Nucl. Phys. {\bf B#1} #2 (#3)} %
\newcommand{\npbps}[3]{Nucl. Phys. B (Proc. Suppl.)  {\bf #1} #2 (#3)}
\newcommand{\zp}[3]{Z.~Phys. C{\bf#1} #2 (#3)}
\newcommand{\epj}[3]{Eur. Phys. J. {\bf C#1} #2 (#3)}
\newcommand{\mpla}[3]{Mod. Phys. Lett. {\bf A#1} #2 (#3)} %
\newcommand{\apj}[3]{ Astrophys. J.\/ {\bf #1} #2 (#3)} %
\newcommand{\jhep}[2]{{Jour. High Energy Phys.\/} {\bf #1} (#2) }%
\newcommand{\astropp}[3]{Astropart. Phys. {\bf #1} #2 (#3)} %
\newcommand{\ib}[3]{{ ibid.\/} {\bf #1} #2 (#3)} %
\newcommand{\nat}[3]{Nature (London) {\bf #1} #2 (#3)} %
\newcommand{\app}[3]{{ Acta Phys. Polon.  B\/}{\bf #1} #2 (#3)}%
\newcommand{\nuovocim}[3]{Nuovo Cim. {\bf C#1} #2 (#3)} %
\newcommand{\yadfiz}[4]{Yad. Fiz. {\bf #1} #2 (#3); 
Phys. {\bf #1} #3 (#4)]} %
\newcommand{\jetp}[6]{{Zh. Eksp. Teor. Fiz.\/} {\bf #1} (#3) #2; {JETP
} {\bf #4} (#6) #5}%
\newcommand{\philt}[3]{Phil. Trans. Roy. Soc. London A {\bf #1} #2
        (#3)}                                                          %
\newcommand{\hepph}[1]{hep--ph/#1}           %
\newcommand{\hepex}[1]{hep--ex/#1}           %
\newcommand{\astro}[1]{(astro--ph/#1)}         %


\begin{thebibliography}{20}

\bibitem{Choi:2001ww}
S.~Y.~Choi, J.~Kalinowski, G.~Moortgat-Pick and P.~M.~Zerwas,
Eur.\ Phys.\ J.\ C {\bf 22}, 563 (2001) [Addendum-ibid.\ C {\bf 23},
769 (2002)] [arXiv:hep-ph/0108117];
J.~Kalinowski, G.~Moortgat-Pick~[arXiv:\hepph{0202083}].

\bibitem{pnp} 
P.~N.~Pandita, Phys. Rev. {\bf D~53}, 566 (1996);
P.~N.~Pandita~[arXiv: hep-ph/9701411].

\bibitem{neutgen} A.~Bartl, H.~Fraas, W.~Majerotto, Nucl. Phys. {\bf
B278}, 1 (1986); J.F.~Gunion, H.E.~Haber, Phys. Rev {\bf D37},2515
(1988); A.~Bartl, H.~Fraas, W.~Majerotto, N.~Oshimo, Phys. Rev. {\bf
D40}, 1594 (1989).

\bibitem{neutdec} S.~Ambrosanio, B.~Mele, Phys. Rev {\bf D55}, 1399 (1997).

\bibitem{neutDM} D.~Majumdar, J. Phys. {\bf G28}, 2747 (2002); Y.~Gyun,
T.~Nihei, L.~Roszkowski, R.~de Austri, JHEP {\bf 0212}, 034 (2002).

\bibitem{Berezinsky:1995cj} V.~Berezinsky, A.~Bottino, J.~R.~Ellis,
N.~Fornengo, G.~Mignola and S.~Scopel,
Astropart.\ Phys.\ {\bf 5}, 1 (1996) [arXiv:hep-ph/9508249].

\bibitem{prs} S.~Pokorski, J.~Rosiek, C.A.~Savoy, Nucl. Phys. {\bf
B570}, 81 (2000).

\bibitem{Choi:1998ei} S.~Y.~Choi, A.~Djouadi, H.~S.~Song and
P.~M.~Zerwas,
Eur.\ Phys.\ J.\ C {\bf 8}, 669 (1999) [arXiv:hep-ph/9812236];
S.Y.~Choi et al.,Eur.\ Phys.\ J.\ C {\bf 14}, 535
(2000)[arXiv:hep-ph/0002033] .

\bibitem{Allanach:2001kg} B.~C.~Allanach, ``SOFTSUSY: A C++ program
for calculating supersymmetric spectra,'' Comput.\ Phys.\ Commun.\
{\bf 143}, 305 (2002) [arXiv:hep-ph/0104145].

\bibitem{nilles} H.~P.~Nilles, Phys. Rep. {\bf 110}, 1~(1984); Pran
Nath, R.~Arnowitt, and A.~H.~Chamseddine, in {\it Supersymemtry,
Supergravity and Perturbative QCD}, edited by P.~Roy and V.~Singh
(Springer, Heidelberg, 1984).

\bibitem{ggw} T.~Gherghetta, G.~F.~Giudice and J.~D.~Wells,
Nucl.Phys. {\bf B559}, 27 (1999) [arXiv:hep-ph/9904378].

\bibitem{Pierce:1996zz}
D.~M.~Pierce, J.~A.~Bagger, K.~T.~Matchev and R.~J.~Zhang,
Nucl.\ Phys.\ B {\bf 491}, 3 (1997) [arXiv:hep-ph/9606211].

\bibitem{pdg} K.~Hagiwara et al., Particle Data Group, Phys. Rev. {\bf
D66}, 010001 (2002).

\bibitem{giudice} G.~F.~Giudice, M.~A.~Luty, H.~Murayama, and
R.~Rattazzi, \jhep{9812:027}{1998}~[arXiv:hep-ph/9810442].

\bibitem{randall} L.~Randall and R.~Sundrum~[arXiv:hep-ph/9810155].

\bibitem{pr} A.~Pomarol and R.~Rattazzi, JHEP~{\bf 9905}, 013 (1999)~
[arXiv:hep-ph/9903448].

\bibitem{fix} E.~Katz, Y.~Shadmi, Y.~Shirman, \jhep{9908: 015}{1999};
R.~Rattazzi, A.~Strumia, J.~D.~Wells, \np{576}{3}{2000}; Z.~Chacko,
M.~Luty, E.~Pont\'{o}n, Y.~Shadmi, Y.~Shirman, \prd{64}{055009}{2001};
Z.~Chacko, M.~A.~Luty, I.~Maksymsk, E.~Pont\'{o}n,
\jhep{0004:001}{2000}; I.~Jack, D.~R.~T.~Jones, \np{482}{167}{2000};
M.~Carena, K.~Huitu, T.~Kobayashi, \np{592}{164}{2001}.

\bibitem{hlp} For a recent study, see K.~Huitu, J.~Laamanen and
P.~N.~Pandita, Phys. Rev. {\bf D65}, 115003
(2002)~[arXiv:hep-ph/0203186]; K.~Huitu, J.~Laamanen and
P.~N.~Pandita~[arXiv:hep-ph/0303067].

\end{thebibliography}
\end{document}